\documentstyle[aps,multicol,epsfig]{revtex}

\begin{document}

\title{Off equilibrium dynamics of the Frustrated Ising Lattice Gas}
\author{Daniel A. Stariolo$^{1,2}$ and Jeferson J. Arenzon$^2$}
\address{$^1$Departamento de F{\'\i}sica, Universidade Federal de
Vi\c cosa\\
36570-000 Vi\c cosa MG, Brazil}
\address{$^2$Instituto de F{\'\i}sica, Universidade Federal do Rio Grande do Sul\\ 
CP 15051, 91501-970 Porto Alegre RS, Brazil \\ E-mails: {\tt
stariolo@mail.ufv.br, arenzon@if.ufrgs.br}}
\date{\today}
\maketitle

\begin{abstract}
We study by means of Monte Carlo simulations the off equilibrium properties of a
model glass, the Frustrated Ising Lattice Gas (FILG) in three
dimensions. We have computed
typical two times quantities, like density-density autocorrelations and 
the autocorrelation of internal degrees of freedom. We
find an aging scenario particularly interesting in the case of the density
autocorrelations in real space which is very reminiscent of spin glass 
phenomenology. While this model captures the essential features of structural 
glass dynamics, its analogy with spin glasses may bring the possibility of 
its complete description using the tools developed in spin glass theory. 
\end{abstract}

%\thispagestyle{empty}
%\newpage

\begin{multicols}{2}
\narrowtext

Much effort is being currently devoted to reach a reasonable theoretical
understanding
of structural glass physics. While the huge amount of experimental work 
available shows an extremely rich phenomenology, a succesfull description in terms
of microscopic models is still lacking. The theoretical models available are
mainly phenomenological, basically focusing the non-Arrhenius relaxation
in the so called fragile glasses~\cite{angell}. Among the most succesfull
ones are the {\it free volume model}~\cite{cohen} and the {\it entropy
model}~\cite{gibbs} in which the relevant variables for the description of a glass
transition from a supercooled liquid phase are respectively volume and entropy.
While the last model predicts the existence of a thermodynamic
second order transition at a well defined temperature, it is extremely difficult to
obtain evidences of it as it is in practice experimentally inaccessible.
Another succesfull approach has been the {\it Mode Coupling
Theory} of G{\"o}tze and S{j{\"o}}gren~\cite{gotze}. This dynamical approach is
qualitatively correct in
predicting the behavior of time correlations and responses in supercooled liquids.
Recently is has been shown to correspond to the high temperature limit of a dynamical
theory of spin glasses~\cite{bouchaud}. This raised the interesting possibility of
describing the structural glass physics by exploiting the analogy with
some spin glasses in which the transition is discontinuous
as originally suggested by Kirkpatrick et al.~\cite{kirkpatrick}.
Moreover, some finite dimensional models have been proposed
whose mean field limit would provide such a transition
\cite{FrPa98}.
Nevertheless, although spin and structural glasses have some basic features in
common, like the characteristic slow dynamics, 
an essential difference is the absence
(presence) of quenched disorder in structural glasses (spin glasses). 
Unlike in spin glasses, the nature of the glass transition may be 
purely dynamical in origin, without an underlying thermodynamic 
transition. The existence of a growing correlation length,
for example, has not been considered up to very recently in structural glass 
models~\cite{giorgio}. The absence of a simple microscopic model turns
these questions 
very difficult to be answered and much of the present knowledge
comes from computer 
simulations of Lennard-Jones systems~\cite{kob} or from simple
kinetically constrained models~\cite{andersen,peliti,FoRi96,GrPiGr97}.
These, in particular, have dynamically self-induced frustration effects
obtained by restricting the possible Monte Carlo movements and
reproduce reasonably well the glassy phenomenology. However, although
having
the advantage of being lattice models, it is not obvious how to
relate the {\it ad hoc} kinetic rules with the underlying physics.
%Analogously, other models without disorder with glassy behavior 
%have been introduced in contexts as diverse
%as glasses~\cite{nodis}, Josephson junctions~\cite{josephson} and 
%sandpiles~\cite{sand}.

A possible candidate to fill in this gap is the Frustrated Ising Lattice Gas model 
(FILG)~\cite{fp}, a Hamiltonian lattice gas in which the presence of
internal degrees of freedom subjected to quenched disorder mimics the geometric
frustration that slows down the motion of the molecules as the system is cooled or
compressed. The introduction of internal degrees of freedom is responsible for the
slowing down of the diffusional dynamics of the particles and the necessity of 
considering them in order to describe the essence of glass
physics has recently been addressed by Tanaka~\cite{tanaka}. The thermodynamic
properties of the FILG in 3D as well as its equilibrium dynamics have been studied 
by Nicodemi and Coniglio~\cite{nicodemi}. The model shows many
glass properties, including dependence on the
cooling (or compression) rate,
stretched exponential behavior in correlation functions, a dynamical singularity
in which the diffusion constant goes to zero, the breakdown of the Stokes-Einstein
relation along with anomalous diffusion at intermediate times. 
As in many glasses no singularities in the linear susceptibilities are
observed, in particular the compressibility is continuous everywhere.
Scarpetta {\it et
al.}~\cite{SFP} studied an equivalent, non local version of the FILG (the
Site Frustrated Percolation, SFP) in 2D, finding 
a behavior similar to the 3D
FILG, the main difference being the (Arrhenius) dynamical singularity occurring at zero
temperature. This model also seems to relate the glass transition with 
a percolation-type transition as evidenced by the onset of several precursor
phenomena \cite{SFP}.

Non equilibrium phenomena, aging being one example, are widespread
on a great variety of systems, including polymers, granular
materials and ferromagnetic coarsening~\cite{struik,leticia,bray} and
 appears in the relaxation dynamics of both spin and structural
glasses, either theoretically and experimentally.
In this letter we report results for the off equilibrium dynamics
of the FILG in three dimensions, showing a characteristic aging dynamics
present in some two times quantities and some possible scaling scenarios.
Of particular interest are the
results for the density autocorrelations which suggest the definition of a
non-linear compressibility which may bring information on a possible thermodynamic
transition analog to the magnetic transition in spin glasses.

The FILG is defined by the Hamiltonian:
\begin{equation}
H = -J \sum_{<ij>} (\varepsilon_{ij} \sigma_i \sigma_j - 1)n_i n_j 
- \mu \sum_i n_i .
\end{equation}
There are two kinds of dynamical
variables: the site occupation $n_i=0,1$ ($i=1\ldots N$) and the particles
internal degrees of freedom, $\sigma_i=\pm1$. The usually complex spatial
structure of the molecules of glass forming liquids, which can assume
several spatial orientations, is in part responsible for
the geometric constraints on their mobility. Here we take the simplest case
of two possible orientations,
and the steric effects imposed on a particle by its neighbors
are felt as restrictions on its orientation due to the quenched random
variables $\varepsilon_{ij}=\pm 1$. The key role of the first term of
the Hamiltonian is that when
$J\rightarrow \infty$ (recovering the SFP) no frustrated link can be fully
occupied, implying that any frustrated loop in the lattice
will have a hole and then $\rho<1$ preventing the system from reaching the close
packed configuration.
Finally $\mu$ represents a chemical potential ruling the system
density (at fixed volume) and, by taking $\mu\to\infty$ we recover
the Edwards-Anderson spin glass model. It should also be mentioned that,
after including gravity, the same model successfully describes granular 
materials under vibration~\cite{mimmogranular,arenzon98}, another class
of systems where geometric frustration rules its behavior. 

By increasing $\mu$ the model presents two characteristic
points~\cite{nicodemi,SFP}. 
For $\mu \leq 0.75$ (low density) it shows
liquid like behavior, time correlation functions decay exponentially,
equilibration is quickly achieved and the particles mean squared 
displacement grows linearly with time, a simple diffusion scenario. At
$\mu \approx 0.75$ there is a percolation transition (the
corresponding density being $\rho \approx 0.38$).
%, similar to the random site percolation~\cite{stauffer}).
Dynamically it manifests in the onset of two 
different relaxation regimes in the
correlation functions, a fast exponential
relaxation at short times and a slow relaxation at longer times characterized by
stretched exponentials. 
The diffusion is still linear for long times but the diffusion
coefficient becomes smaller as the density grows. Also, for
a fixed $\mu$, the equilibrium density depends on the cooling rate.
The dynamics becomes slower as the chemical potential grows (or equivalently,
the temperature is lowered) and a second transition is reached for
$\mu \geq 6$. This is a spin glass transition associated with the 
frozen in of the internal degrees of
freedom. Interestingly, at this point a dynamical singularity is also
present, manifested by the vanishing of the diffusion constant.
Besides the qualitative analogies with the physical processes typical of
glass forming liquids, there are a
few points of more fundamental character. The presence of two characteristic
temperatures (or chemical potentials) separating
different dynamical regimes is common with a class of mean field theories of
spin glasses and
also with what is observed in real glass formers. The percolation transition
corresponds to the dynamical transition in mean field $p$-spin or Potts glasses or 
with the
glass transition in the mode coupling theory. In the FILG the dynamical 
signature of
the transition is consequence of the appearance of percolating clusters, 
a geometrical
feature which may be responsible for the slowing down of the dynamics 
in structural glasses. 
The second transition corresponds to the {\it ideal glass transition} at 
which the relaxation times diverge and the diffusion constant goes to zero. 
In the
FILG the structural manifestation of this transition is the presence of a 
{\it frozen}
percolating cluster. The density of the system approaches a critical value
$\rho_c \approx 0.7$, attainable only by an infinitely slow cooling. 
These facts turn the FILG an important model for studying different aspects
of the glass transition.

Different protocols can be envisaged for studying, for example, 
correlations and response functions. We have prepared the model in a non 
equilibrium state, setting the parameters as in~\cite{nicodemi}
with an initial density lower than the critical one. Then
the system is quenched to a super critical chemical potential and dynamic
correlations are recorded. We have used a Monte Carlo dynamics which alternates 
flipping
of the internal degrees of freedom, creation and destruction of particles in a
plane surface (which mimics a compression experiment) as in~\cite{peliti} and 
particle diffusion. The creation-destruction of
particles is fast enough as to destroy all particles at long times if it is allowed
in the bulk.

\begin{figure}%[tbh]
\centerline{\epsfig{file=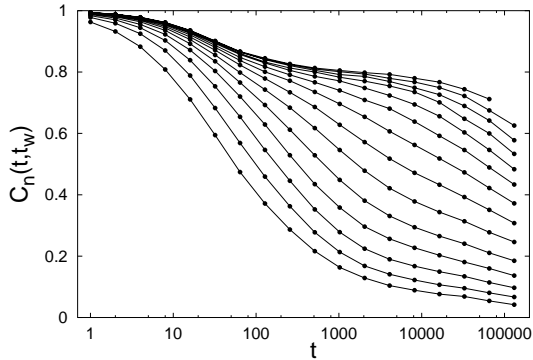,width=5cm,angle=270}}
\caption{Density autocorrelations after a quench to 
$\mu=10$ at $T=1$ and $J=10$ for $L=20$. The waiting times range from
$2^5$ (bottom) to $2^{17}$ (top) and the averages are over 50 samples.}
\label{density}
\end{figure}

Consider the connected two point correlation 
\begin{equation}
c(t,t_w) = \frac{1}{N}\sum_i n_i(t+t_w)n_i(t_w) - \rho(t+t_w) \rho(t_w)
\end{equation}
where the global density at time $t$ is given by
$\rho(t)=N^{-1}\sum_i n_i(t)$. We now define the density
autocorrelations as $C_n(t,t_w) = c(t,t_w)/c(0,t_w)$.
In figure (\ref{density}) the behavior of 
$C_n(t,t_w)$ is shown as a function of $t$ in a semi-log plot after a quench in
chemical potential
to a value $\mu=10$, for waiting times between $2^5$ and $2^{17}$. A
typical
aging scenario is present signalling the slowing down of the dynamics as the
waiting time grows. For the longest waiting times the correlation presents a
rather fast relaxation to a plateau in which the system is in quasi-equilibrium:
the dynamics is stationary and the fluctuation-dissipation relations hold. 
The plateau separates two time 
scales typical of glassy systems: a $\beta$ (fast) relaxation for small time
and an $\alpha$ (slow) relaxation at longer times, corresponding
respectively to the
fast movements of the particle inside the dynamical cages and the large scale,
cooperative process that takes much more time in order to rearrange the cages.
Moreover, in this very long time regime ($t \gg t_w$), the system falls out of equilibrium, the
correlations decay to zero asymptotically and time translational invariance (TTI)
no longer holds with the corresponding violation of the 
fluctuation-dissipation theorem (FDT).

\begin{figure}%[tbh]
\centerline{\epsfig{file=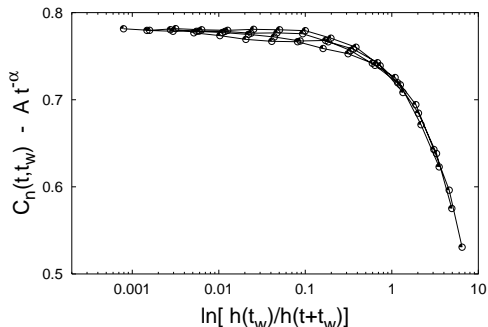,width=4.5cm,angle=270}}
\caption{Scaling plot of the five upper curves of figure (\ref{density}).
The initial five points ($\beta$ decay) were discarded and the parameters
are $A=0.59$, $\alpha=0.46$ and $\mu=0.9$. See text for details.}
\label{C_n_scaling}
\end{figure}

As the $C_n(t,t_w)$ curves have a complex behavior it is possible to scale the curves 
only after a careful analysis of the different time scales present\cite{vincent}. 
It is clear that only for the largest waiting times different
relaxation scales can be observed. We restrict our analysis to the five upper curves
$(t_w=2^{13} \ldots 2^{17})$. In these curves, a fast initial decay is observed up to times
$t \sim 100\;M\!C\!S$, corresponding to the already mentioned $\beta$ decay which we will
not analyze in any more detail. Then a plateau characteristic of quasi-equilibrium
dynamics develops. Its asymptotic value as $t_w\rightarrow \infty$ defines the
ergodicity breaking (or Edwards-Anderson) parameter. For a fixed, large
$t_w$ and $t \gg t_w$, the system begins to
fall out of equilibrium and the correlations decay to zero. But only the very early epochs of
this decay can be observed within the times of our simulations, although we can observe the crossover
between equilibrium and non equilibrium dynamics. Consequently, in order to obtain a good
 scaling for
this last regime we have subtracted from $C_n(t,t_w)$ a stationary contribution of the form:
\begin{equation}
C_n(t,t_w) = C_{\infty} + A t^{-\alpha}, \hspace{1cm}t_w\,\mbox{fixed}\,, t\ll t_w.
\end{equation}
with $C_{\infty}, A$ and $\alpha$ as fit parameters. For the non stationary regime we assumed a
time dependence of the form $h(t_w)/h(t+t_w)$ with $h(x)$ given by\cite{vincent}
$h(x)=\exp\left[(1-\mu)^{-1}(x/\tau)^{1-\mu}\right]$ with $\mu < 1$ and $\tau$ a microscopic
timescale. Note that this form is quite general as one recovers the cases of a simple
$t/t_w$ dependence (full aging) when $\mu=1$ and stationary dynamics when $\mu=0$. The final
scaling is shown in figure (\ref{C_n_scaling}). The scaling obtained is much better
than assuming only full aging or activated dynamics even after including in these cases the
contribution from the stationary decay.

\begin{figure}%[htbp]
\centerline{\epsfig{file=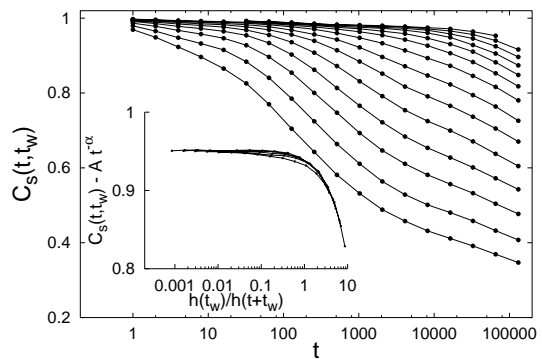,width=5cm,angle=270}}
\caption{Autocorrelation of the internal degrees of freedom after a quench
to $\mu=10$ with $T=1$ and $J=10$ for $L=20$. Waiting times from $2^5$ to $2^{17}$ 
(top to bottom) and average over 50 samples. Inset: 
scaling with $\tau_0=0.5$, $\mu=0.9$, $\alpha=0.091$ and $A=0.057$
(see text).}
\label{spin}
\end{figure}

We have also measured, as shown in figure (\ref{spin}), the autocorrelations 
of the internal degrees of freedom:
\begin{equation}
C_s(t,t_w) = \frac{1}{N}\sum_i s_i(t+t_w)n_i(t+t_w)s_i(t_w)n_i(t_w),
\end{equation}             
The internal degrees of freedom correspond to a diluted Edwards-Anderson (EA) 
spin glass and 
the curves should be compared to the ones of the 3D EA model~\cite{rieger3D}. 
%The dynamics of these ``spin''
%variables is slower than that of the density ones,
%and can be seen that the EA parameter
%in this case will be slightly lower than one meaning that one valley occupies only
%a tiny fraction of phase space. 
%Nevertheless, the system is not completely frozen and
%after a sufficiently long time it escapes from the valley and goes out of equilibrium.
From our knowledge of the 3D EA model one may expect the scaling of the
autocorrelation in the aging regime be of the form 
$t^{-\alpha}\tilde{C}(t/t_w)$. 
Nevertheless we verified that an activated dynamics scaling of the form 
$C_s(t,t_w)\propto \tilde{C}(\log(t)/\log(t_w))$
works better. Moreover, doing an analysis similar to the one for $C_n(t,t_w)$
the scaling can be slightly improved (see fig.(\ref{spin}), inset).

To our knowledge this is the first Hamiltonian lattice model which presents the
essentials of structural glass phenomenology. In particular the results 
for the density autocorrelation are very promising
suggesting that it may be possible to apply spin glass ideas and techniques 
for an analytical investigation of the model\cite{jef}.  The similarity between
the density autocorrelation and the spin glass overlap function suggests the 
introduction of a non-linear compressibility in analogy with the spin glass
susceptibility, which may be a good quantity for studying the possibility of an
underlying phase transition with a growing correlation length in the model.
Another issue that must be studied is the precise form of the violation of FDT
through the so called {\it fluctuation dissipation ratio}~\cite{silvio}.
Also other protocols may be implemented in order to probe the off
equilibrium dynamics of the system. Work in progress, keeping fixed the global
density and performing a quench in temperature, indicates that the phenomenology
is similar to what we have presented here. Moreover, the evaluation of the root mean
square deviation of the particles and the incoherent scattering function (which
is related to the Fourier transform of the density correlations) also show evidence of
slow dynamics and aging. The results of these investigations will be published in a
future paper. We are working also in the $2D$ version of the 
model: here, some differences are expected with respect to $3D$
because the dynamical
singularity only occurs for $T=0$ $(\mu\rightarrow \infty)$ and the relaxation
time diverges with an Arrhenius law. 
Still other issues can also be explored. For instance, to what extent, if any,
the scenario for the $3D$ model presented here changes as one goes to 
infinite range connections. In the mean field version, where both
first and second order transitions show up, it might be possible that
different aging regimes are present~\cite{jef}, and would be 
interesting to know if the FILG
is a finite dimensional version of a model whose mean field limit has 
a discontinuous transition.

This work was partly supported by Brazilian agencies CNPq and FAPEMIG.
We acknowledge M. Sellitto for a careful
reading of the manuscript and J.A.C. Gallas for providing time on his
alpha station.
%Conselho Nacional de Desenvolvimento 
%Cient\ii fico e Tecnol{\'o}gico (CNPq), Brazil.

\end{multicols}
\end{document}